\documentclass[letterpaper,twocolumn,10pt]{article}
\usepackage{usenix2019_v3}
\pdfoutput=1

\usepackage{multirow}
\usepackage{amsmath,amssymb,amsfonts}
\usepackage{algorithmic}
\usepackage{graphicx}
\usepackage{textcomp}
\usepackage[table,xcdraw]{xcolor}
\usepackage{hyperref}
\usepackage{subcaption}
\usepackage[justification=centering]{caption}
\usepackage{bm}
\usepackage[compact]{titlesec}
	\titlespacing{\section}{0pt}{2ex}{1ex}
	\titlespacing{\subsection}{0pt}{1ex}{1ex}
	\titlespacing{\subsubsection}{0pt}{1ex}{0ex}
	\titlespacing{\paragraph}{0pt}{1.25ex plus 1ex minus .2ex}{1em}
\usepackage{enumitem}
\usepackage{titling}
\usepackage{dblfloatfix}

\graphicspath{{./}}

\titleformat*{\section}{\large\bfseries}
\titleformat*{\subsection}{\normalsize\bfseries}
\titleformat*{\subsubsection}{\normalsize\bfseries}
\titleformat*{\paragraph}{\normalsize\bfseries}
\titleformat*{\subparagraph}{\normalsize\bfseries}

\AtBeginDocument{%
	\abovedisplayshortskip=0pt plus 3pt
}

\posttitle{\par\end{center}}

\setlist[itemize]{noitemsep, topsep=0pt}
\setlist[enumerate]{noitemsep, topsep=0pt}

\newcommand\Tstrut{\rule{0pt}{2.6ex}}       
\newcommand\Bstrut{\rule[-0.9ex]{0pt}{0pt}} 
\newcommand{\TBstrut}{\Tstrut\Bstrut} 

\hypersetup{
	colorlinks   = true, 
	urlcolor     = black, 
	urlbordercolor=black, 
	linkcolor    = blue, 
	linkbordercolor = white, 
	citecolor    = blue, 
	citebordercolor = white, 
	breaklinks   = true, 
}
\makeatletter
\Hy@AtBeginDocument{
	\def\@pdfborder{0 0 1} 
	\def\@pdfborderstyle{/S/U/W 0.5} 
}
\makeatother

\usepackage[
backend      = biber,
sorting      = none, 
citestyle    = numeric-comp, 
maxbibnames  = 3, 
mincrossrefs = 1000, 
]{biblatex} 
\bibliography{proceedings.bib,disinfo-papers.bib}

\DeclareSourcemap{
	\maps[datatype=bibtex, overwrite]{
		\map{
			\step[fieldset=editor, null]
			\step[fieldset=location, null]
			\step[fieldset=publisher, null]
			\step[fieldset=isbn, null]
			\step[fieldset=issn, null]
			\step[fieldset=pages, null]
		}
	}
}

\date{}

\title{\Large \bf Adapting Security Warnings to Counter Online Disinformation}

\author{
	{\rm Ben Kaiser}\\ Princeton University
	\and
	{\rm Jerry Wei}\\ Princeton University
	\and
	{\rm Eli Lucherini}\\ Princeton University
	\and
	{\rm Kevin Lee}\\ Princeton University
	\and
	{\rm J.\ Nathan Matias}\\ Cornell University
	\and
	{\rm Jonathan Mayer}\\ Princeton University
}

\begin{document}

\maketitle

\begin{abstract}
  Disinformation is proliferating on the internet, and platforms are responding by attaching warnings to content. There is little evidence, however, that these warnings help users identify or avoid disinformation. In this work, we adapt methods and results from the information security warning literature in order to design and evaluate effective disinformation warnings.

  In an initial laboratory study, we used a simulated search task to examine contextual and interstitial disinformation warning designs. We found that users routinely ignore contextual warnings, but users notice interstitial warnings---and respond by seeking information from alternative sources.

  We then conducted a follow-on crowdworker study with eight interstitial warning designs. We confirmed a significant impact on user information-seeking behavior, and we found that a warning's design could effectively inform users or convey a risk of harm. We also found, however, that neither user comprehension nor fear of harm moderated behavioral effects.

	Our work provides evidence that disinformation warnings can---when designed well---help users identify and avoid disinformation. We show a path forward for designing effective warnings, and we contribute repeatable methods for evaluating behavioral effects. We also surface a possible dilemma: disinformation warnings might be able to inform users and guide behavior, but the behavioral effects might result from user experience friction, not informed decision making.
\end{abstract}

\thispagestyle{empty}
\pagestyle{empty}

\vspace{-1mm}
\section{Introduction}
\label{sec:introduction}


Disinformation is spreading widely on the internet, often propelled by political motives~\cite{bradshaw2019global,martin2019recent}.
Platforms are responding by attaching warnings to disinformation content, in order to inform users and guide their actions.
Facebook implemented disinformation warnings as early as December 2016~\cite{mosseri2016addressing}, and Google~\cite{kosslyn2017fact}, Bing~\cite{bing2017fact}, and Twitter~\cite{alba2020twitter} have adopted similar content notices.
There has been substantial public debate about the propriety of disinformation warnings, especially after Twitter began labeling tweets by U.S. President Donald Trump in May 2020~\cite{hatmaker2020twitter}. But recent studies provide scant evidence that these warnings can meaningfully influence user beliefs or behaviors, and it is an open question whether warnings are promising or futile for combating disinformation.

Security researchers faced a similar challenge over a decade ago, when studies showed that warnings for malware, phishing, and other online threats broadly failed to protect users~\cite{wu2006security, egelman2008youve}.
After a series of iterative, multi-method studies~\cite{akhawe2013alice,sunshine2009crying,harbach2013sorry,felt2014experimenting,felt2015improving,weinberger2016week,malkin2017personalized,reeder2018experience,egelman2013importance,almuhimedi2014your,bravolillo2014revisiting,acer2017where}, security warnings now reliably inform user security decisions and help users avoid harmful and inauthentic content~\cite{akhawe2013alice,reeder2018experience}.
In this work, we adapt methods and results from the information security warning literature in order to design and evaluate effective disinformation warnings.

A key finding from security research that we adapt to disinformation is that \textit{contextual} warnings, which do not interrupt the user or compel action, are far less effective at changing behavior than \textit{interstitial} warnings, which interrupt the user and require interaction~\cite{wu2006security,egelman2008youve,reeder2018experience}.
Our work is, to our knowledge, the first to evaluate interstitial disinformation warnings.

Another relevant contribution from the security literature is a set of rigorous qualitative and quantitative methods for evaluating warnings, including structured models, realistic guided tasks, user interviews, and field studies (e.g.,~\cite{egelman2013importance,sunshine2009crying,felt2014experimenting,weinberger2016week,malkin2017personalized,reeder2018experience}).
Our work adapts these methods to empirically examine contextual and interstitial disinformation warnings.

Across two studies, we use qualitative approaches (think-aloud exercises, interviews, and inductive coding) to understand user perceptions of disinformation warnings, as well as quantitative measures of the warnings' effects on user behavior.
We consider the following research questions:

\begin{enumerate}
	\item After encountering contextual and interstitial disinformation warnings, how often do users change their behavior by opting for alternative sources of information?
	\item Why do some users choose not to change their behaviors after encountering contextual and interstitial disinformation warnings?
	\item Can interstitial warnings that are highly informative effectively change user behavior?
	\item Can interstitial warnings that are highly threatening effectively change user behavior?
\end{enumerate}

We first conducted a laboratory experiment ($n = 40$) in which participants searched for specific facts on Google and encountered an interstitial or contextual disinformation warning for certain search results (Section~\ref{sec:labstudy}).
The interstitial warning was substantially more effective at changing user behavior than the contextual warning, in large part because users did not notice or comprehend the more subtle contextual warning.
In post-task interviews, participants described two reasons for the interstitial warning's strong behavioral effect: the \textit{informativeness} of the warning's messaging and the \textit{risk of harm} conveyed by the warning's threatening design.

We then conducted a follow-on crowdworker study ($n = 238$), examining eight interstitial warning designs (Section~\ref{sec:mturkstudy}). We confirmed the strong behavioral effects of interstitial warnings.
We also found, however, that neither user comprehension nor perceived risk of harm appeared to moderate those effects.

Our results provide evidence that interstitial disinformation warnings can both inform users and guide user behavior. We demonstrate scalable and repeatable methods for measuring warning effectiveness and testing theories of effect. We also surface a possible dilemma: the behavioral effects of disinformation warnings may be attributable to user experience friction, rather than informed decision making. Our work highlights a path forward for designing effective warnings, and we close by encouraging iterative research and improvement for disinformation warnings---just like the information security community has successfully done for security warnings.

\vspace{-1mm}
\section{Background and Related Work}
\label{sec:background}
Disinformation research is dispersed across academic disciplines.\footnote{We use the term ``disinformation'' here and throughout this work, because our studies focus on warning users about intentionally misleading websites. We believe that our results generalize to misleading content regardless of intent (i.e., ``misinformation''), because laboratory participants did not identify a website's motive as a salient concern and we find that interstitial warning behavioral effects are not significantly related to warning content.} Recent work has predominantly focused on measurement (e.g., of content, campaigns, or user interactions)~\cite{delvicario2016spreading,allcott2017social,allcott2019trends,burger2019reach,guess2020exposure,zubiaga2016analysing,howard2017social,shao2018spread,vosoughi2018spread,allcott2019trends,kumar2016disinformation,bush2019bing,fourney2016geographic,shao2018anatomy,badawy2019falls,bovet2019influence,grinberg2019fake,guess2020exposure,ferrara2017disinformation,cantarella2019does,starbird2018ecosystem}, or on developing automated detection methods~\cite{vlachos2015identification,thorne2017extensible,hassan2017claimbuster,thorne2018automated,thorne2018fever,rashkin2017truth,potthast2018stylometric,horne2019this,arXiv:2003.07684,yang2012automatic,wu2015false,zhao2015enquiring,ma2017detect,baly2018predicting,popat2018where}.

In this section, we begin with background on disinformation websites, which are the targets for the warnings in our studies. We then discuss related work on the effects of and responses to disinformation. Finally, we describe the security warnings literature, which is the inspiration for this work.

\subsection{Disinformation Websites}
Disinformation campaigns are often multimodal, exploiting many different social and media channels at once~\cite{wardle2017information}.
These campaigns use websites as an important tool to host content for distribution across platforms and generate ad revenue~\cite{howard2017social,bradshaw2018junk,bengani2019hundreds,burger2019reach,calishain2019junk,kasprak2019hiding}.
Disinformation websites are designed to intentionally deceive users into believing that they are legitimate news outlets.\footnote{
	 These websites are sometimes termed ``fake news'' or ``junk news''  in related work (e.g.,~\cite{lazer18science, bradshaw2018junk,yin2018friendly}).
}
Our work examines whether warnings can counter this deception and help users contextualize or avoid disinformation websites.

\subsection{Effects of Disinformation}
\label{subsec:effectsofdisinfo}
Disinformation campaigns hijack the heuristics that users rely on to make accurate judgments about the truthfulness of information~\cite{lewandowsky2012misinformation}.
For example, disinformation campaigns often mimic credibility indicators from real news sources~\cite{tandoc2018defining} or use social media bots to create the appearance of support~\cite{shao2018spread}.

Misperceptions that individuals hold after consuming disinformation are difficult to dispel~\cite{lewandowsky2012misinformation}.
Collectively, a misinformed populace may make social and political decisions that are not in the society's best interests~\cite{kuklinski2000misinformation} (e.g., failing to mitigate climate change~\cite{maertens2020combatting}).
Influencing policy in this way---by shaping public perception and creating division---is a goal of many campaigns, especially by state-level actors~\cite{nemr2019weapons}.

Presenting a warning before exposure to disinformation can prevent harmful effects in several ways.
Warnings can induce skepticism, so that users are less likely to take disinformation at face value~\cite{greene1982inducing}.
Warnings can also make users more susceptible to corrections~\cite{ecker2010explicit,lewandowsky2012misinformation}.
Finally, warnings may cause users to not read the disinformation at all.

\subsection{Responses to Disinformation}
\label{subsec:respondingtodisinformation}
There are three main types of responses to disinformation that platforms and researchers have considered~\cite{goldman2019content}.
The first is deranking disinformation by changing recommendation algorithms~\cite{taylor2018industry}. Academics have studied this approach in simulated models of social networks~\cite{kimura2008minimizing,budak2011limiting,nguyen2012containment,wang2013negative,farajtabar2017mitigation,li2019optimization}, although not in realistic settings or with real users.
Second, platforms have repeatedly removed disinformation content and banned accounts that promote disinformation~\cite{twitter2019disclosing,facebook2018removing,clegg2020combatting}.
Neither platforms nor researchers have established evidence on the effects of these takedowns.
Finally, platforms have added warnings and similar forms of context to posts~\cite{smith2018designing,kosslyn2017fact,bing2017fact,alba2020twitter, singh2020whatsapp}.

We note that there are important speech distinctions between these responses. When a platform removes content, it unilaterally makes speech less accessible to users. When a platform deranks content, it leaves the content available, but it unilaterally curtails speech distribution and discoverability. The potential promise of disinformation warnings is that they respond to problematic speech with counterspeech: platforms inform and protect users, without making unilateral decisions about content availability, distribution, or discoverability. As we discuss in Section~\ref{subsec:mturkdisc}, our work poses a possible dilemma for disinformation warnings as speech regulation: warnings can inform users and guide user behavior, but the behavioral effects may not be attributable to informed decision making.

\paragraph{Fact Check Warnings}
The most well-studied disinformation warnings are contextual labels indicating a story has been ``disputed'' or ``rated false'' by fact checkers~\cite{pennycook2018prior, ross2018fake, gao2018label, clayton2019real, pennycook2019implied, seo2019trust}.
These labels constituted Facebook's first major effort to counter disinformation~\cite{smith2018designing}, and Google~\cite{kosslyn2017fact}, Bing~\cite{bing2017fact}, and Twitter~\cite{alba2020twitter} have taken similar approaches.
Facebook eventually discontinued use of ``disputed'' warnings after determining based on internal studies that the warnings were of limited utility~\cite{smith2018designing}.
More recently Facebook, Instagram, and Twitter all deployed new warning formats, including interstitials~\cite{rosen2019helping,collins2020twitter}.

Some studies of fact check warnings reported no significant effects on participant perceptions of disinformation~\cite{ross2018fake,gao2018label}, while others found moderate effects under certain conditions~\cite{seo2019trust,pennycook2018prior,pennycook2019implied,mena2019cleaning,moravec2020appealing}.
Pennycook et al. found that fact check warnings caused participants to rate disinformation as less accurate after repeated warning exposure, but not with a single exposure~\cite{pennycook2018prior}.
Another study by Pennycook et al. identified a counterproductive \textit{implied truth effect}: attaching a fact check warning to some headlines caused participants to perceive other headlines as more accurate~\cite{pennycook2019implied}.
Seo et al. found that fact check warnings caused participants to perceive stories as less accurate, but the effect did not persist when participants encountered the same stories later~\cite{seo2019trust}.
Mena found that fact check warnings had small negative effects on perceived credibility of news content on social media and self-reported likelihood of sharing~\cite{mena2019cleaning}.
Finally, Moravec et al. examined how fact check warnings can induce instinctual cognitive responses in users and cause users to thoughtfully incorporate new information into their decision making; a warning design that combined both mechanisms showed a moderate effect on social media post believability~\cite{moravec2020appealing}.

\paragraph{Related Links}
Bode and Vraga examined the effects of providing related links to alternative, credible sources of information alongside misinformation~\cite{bode2015related}. The study found that when related links debunked misinformation, participants who initially believed the disinformation showed a limited tendency toward corrected beliefs.
Facebook, Google Search, and Bing all currently use related links warning designs.

\paragraph{Highlighting Falsehoods} Garrett et al. tested a two-part warning, where participants were first informed that a fact-checker had identified factual errors in a story, then those errors were highlighted in the body of the story~\cite{garrett2013promise}.
Among users already predisposed to reject the misinformation, this treatment significantly increased accuracy of beliefs, but it had no effect among users inclined to believe the misinformation.

\paragraph{Methods of Prior Work} In all of these studies, participants were presented with screenshots of simulated social media posts, then were posed survey questions such as how truthful they thought the posts were and whether they would consider sharing the posts on social media.
These methods can inform theories about how users will respond in real-world settings, but generalizations are tenuous because the methods involve highly artificial tasks and self-reported predictions about behavior. As we discuss below, security research has found that in order to measure realistic responses to warnings, it is important to design experimental tasks that involve realistic systems, realistic risks, and measurement of actual participant behavior~\cite{schechter2007emperor,egelman2008youve,sunshine2009crying,sotirakopoulos2011challenges}.

\subsection{Security Warnings}
\label{subsec:securitywarnings}
Effective warnings are essential for security, because there are certain security decisions that systems cannot consistently make on behalf of users. Adversaries deliberately exploit judgment errors associated with these human-in-the-loop security decisions~\cite{cranor2008framework}.
Early studies of security warnings found that the warning formats that were currently in use generally failed to protect users from online risks~\cite{wu2006security,schechter2007emperor}.
Modern warnings, by contrast, are extremely effective: a recent study of over 25 million browser warning impressions showed that the warnings protected users from malware and phishing websites around 75-90\% of the time~\cite{akhawe2013alice}.
The immense progress in security warning effectiveness is due to numerous, rigorous studies that for over a decade have tested varied warning designs using diverse experimental methods and analytic lenses.

The primary methods of early security warning studies were laboratory experiments involving supervised tasks, user interviews, and surveys~\cite{wu2006security,schechter2007emperor,egelman2008youve, sunshine2009crying, sotirakopoulos2011challenges, egelman2013importance,bravolillo2014revisiting}.
These studies examined users' beliefs and decision-making processes, in part by using structured models from warning science literature to identify reasons that warnings failed to change user behaviors.
Security researchers typically used  the Communication–Human Information Processing (C-HIP) model, which describes five information processing stages that users undergo when receiving a warning.
Users must \textit{notice} the warning, \textit{comprehend} the warning's meaning, \textit{believe} the warning, be \textit{motivated} to heed the warning, and finally, \textit{behave} in the manner intended by the warning issuer~\cite{conzola2001communication}.
By determining the stage of the C-HIP model at which information processing was failing, researchers learned how to modify warning designs to increase the strength of the desired effect~\cite{egelman2008youve,sunshine2009crying}.

\paragraph{Limitations}
It can be difficult to cause users to perceive realistic risk in a laboratory, requiring the use of deception and thoughtful experimental design~\cite{schechter2007emperor,sotirakopoulos2011challenges}.
Laboratory studies must also address the challenge that participants may be more likely to disregard warnings that hinder task completion~\cite{sunshine2009crying,sotirakopoulos2011challenges}.
Later research overcame these limitations using field studies, which measure users' reactions to warnings at scale in realistic environments~\cite{akhawe2013alice,almuhimedi2014your,felt2014experimenting,felt2015improving,weinberger2016week,reeder2018experience}.

This body of work has definitively established that actively interrupting a user's workflow with a warning is more effective at preventing risky behaviors than passively showing the warning.
Wu et al. compared popup warnings to toolbar icons and found that the popups caused users to behave with significantly more caution~\cite{wu2006security}; Egelman et al. observed that 79\% of participants chose not to click through interstitial warnings compared to 13\% for contextual warnings~\cite{egelman2008youve}.
As a result, interstitials and other forms of active security warnings have become standard in all major browsers~\cite{reeder2018experience}.

Several studies compared multiple warning designs and found that clear messages and use of visual cues can improve comprehension and adherence~\cite{sunshine2009crying,harbach2013sorry,felt2014experimenting,felt2015improving}.
Personalizing messages based on user-specific psychological factors has not, however, shown a significant effect on adherence~\cite{malkin2017personalized}.

\paragraph{Limits of Analogizing to Disinformation Warnings} The goals of security and disinformation warnings are not identical, so to study disinformation warnings, we must adapt---not simply reuse---the findings and methods from security warning research.
Security warnings protect users from harms that are typically individualized, irreversible, and otherwise difficult for users to avoid themselves.
The risks of disinformation, by contrast, are usually more collective and diffuse (see Section~\ref{subsec:effectsofdisinfo}) and reversible (e.g., by receiving accurate information). Moreover, a user who encounters disinformation may be readily capable of protecting themselves from the risk (e.g., if they are media literate). As noted earlier, disinformation warnings also have speech implications that are distinct from security warnings. The differences between the security and disinformation problem domains motivate us to emphasize designs that inform users throughout our work.

\vspace{-2mm}
\section{Laboratory Study}
\label{sec:labstudy}

We began with a laboratory study designed to examine how participants would process and react to contextual and interstitial disinformation warnings when searching for information.
The search engine context is conducive to studying behavioral effects because participants have a concrete goal (finding a piece of information) and multiple pathways to achieve that goal (different search results and information sources).

We posed three research questions:

\begin{itemize}[wide]
	\item[] \label{item:labRQ1} \textbf{RQ1: In encounters with contextual and interstitial disinformation warnings, do users notice the warnings?} Prior studies of contextual warnings note that one reason effect sizes are low or insignificant is that participants fail to notice the warnings. Effective warnings must attract user attention through conspicuous design or prominent placement. 
	\item[] \label{item:labRQ2} \textbf{RQ2: When users notice contextual and interstitial warnings, do they understand that the warnings have to do with disinformation?} If a user misunderstands a warning, they may drop it from further cognitive consideration or respond in unintended ways that could increase risk.
	\item[] \label{item:labRQ3} \textbf{RQ3: When users encounter and comprehend contextual and interstitial disinformation warnings, do they change their behaviors in the intended way by opting for alternative sources of information?} This is an important outcome of warning exposure, which we aim to measure as described in Section~\ref{subsec:lab-data}.
\end{itemize}

\begin{figure}
\framebox{\includegraphics[width=\linewidth]{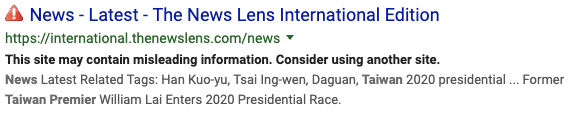}}
\caption{Search results pages displayed contextual warnings.}
\label{fig:contextual}
\end{figure}

\vspace{-1mm}
\subsection{Warning Designs}
We adapted contextual and interstitial disinformation warnings from modern security warnings used by Google.
Google's warnings are well studied~\cite{akhawe2013alice,almuhimedi2014your,felt2015improving,felt2016rethinking,weinberger2016week,acer2017where,reeder2018experience} and widely deployed, making them a useful template to design warnings that participants will believe are real.

We developed our contextual warning (Figure~\ref{fig:contextual}) based on a warning for compromised websites that Google displays in search results.
We changed the color of the text from hyperlink blue to bold black to indicate that the text could not be clicked. We also added a red and white exclamation icon next to the search result to make the warning more noticeable.

\begin{figure}
	\framebox{\includegraphics[width=\linewidth]{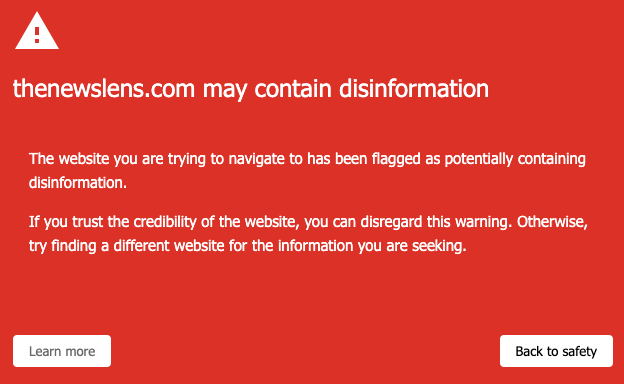}}
	\caption{Participants encountered interstitial warnings after clicking search results.}
	\label{fig:interstitial}
\end{figure}

We adapted our interstitial warning (Figure~\ref{fig:interstitial}) from Google Chrome's warning page for malware.
We modified the text to reference disinformation and changed the ``Details'' button to ``Learn more.''
Clicking ``Learn more'' revealed a message explaining that an automated system had flagged the site as disinformation and a ``Continue'' button that allowed the user to bypass the warning and continue to their selected page.

\subsection{Study Design}
In a laboratory setting, each participant completed a think-aloud role-playing task followed by an interview.
By observing the participant during the task, we could tell if they noticed the warnings (\hyperref[item:labRQ1]{RQ1}) and altered their behavior in response (\hyperref[item:labRQ3]{RQ3}).
Using the interviews, we could confirm whether the participant noticed the warnings, ask whether they comprehended the warnings (\hyperref[item:labRQ2]{RQ2}), and seek additional insights into how the participant processed the warnings.

\paragraph{Role-Playing Task}
The participant assumed the persona of an academic researcher trying to find answers to four questions using Google search.
As our subjects were students, we believed this persona would be comfortable and aid with task immersion.
For each question, we provided multiple sources of information and attached a warning to just one source so that the participant did not have to click through the warning to complete the task.\footnote{Studies of security warnings have shown that this ``task focus effect'' can bias participants' behavior~\cite{sunshine2009crying,sotirakopoulos2011challenges}.}
Unknown to the participant, two questions were control rounds with no warnings and two were treatment rounds where warnings were inserted via a browser extension.
We assigned participants in equal numbers to receive either contextual or interstitial warnings in both treatment rounds.

\paragraph{Search Tasks}
We designed the search tasks (shown in Table~\ref{tbl:laboutcomes}) to cover simple facts that could be easily found with a single Google search.
For the treatment tasks, we selected facts specific to non-U.S. countries and covered by little-known, non-U.S. news sources in order to satisfy three additional design goals.
First, the facts were related to current events due to the study's focus on news disinformation.
Second, so that participants could choose between multiple sources, each fact was publicly reported by at least two English-language news websites.
Third, we aimed to select facts and websites that participants were not likely to be familiar with so as to avoid participants having preconceived biases about the information or the credibility of the sources.

\subsection{Study Procedures}
We explained the task and role to each participant at the beginning of their session.
We asked the participant to behave as if they were using their own computer and to narrate their thoughts and actions as they performed the task.
We framed the study as examining research and search engine usage behaviors to avoid priming participants to think about disinformation or warnings.

Participants began the task on the Google homepage.
We informed participants that they could use either of two specific websites to find a particular piece of information, and that they should start with the first website and could switch to the second for any reason.\footnote{This design directly parallels an evaluation of security warnings conducted by Sunshine et al.~\cite{sunshine2009crying}.}
Control rounds occurred first and third and treatment rounds occurred second and fourth, with the question order randomized within those sets.

We did not prescribe specific search queries to use, but most participants used a similar format: a set of terms relevant to the fact they needed to find combined with the name of the website on which they wanted to find the information.
The participant would enter the query, navigate through results to find the requested information, and verbally provide an answer to the researcher.
We would then instruct them to return to the Google homepage to begin the next round.

\subsection{Data Collection}
\label{subsec:lab-data}
We took notes during each session to record how the participant selected search results, browsed the websites to seek information, reacted to warnings, and described their thoughts and actions.
We also computed two metrics for each warning: a \textit{clickthrough rate} (CTR) and an \textit{alternative visit rate} (AVR).

\paragraph{Clickthrough Rate}
CTR is a commonly used metric in studies of security warnings.
It measures the proportion of warning encounters where a participant dismisses the warning and proceeds, instead of going back.
For contextual warnings, we recorded a clickthrough if the participant clicked a result that had an attached warning and a non-clickthrough if they chose a different result or changed the search query to use the second suggested website.
For interstitial warnings, we recorded a clickthrough if the participant clicked ``Learn more'' and then bypassed the warning.
If the participant clicked ``Back to safety'' or used the browser back button to go back to the search results, we recorded a non-clickthrough.

\paragraph{Alternative Visit Rate}
We also recorded whether participants visited an alternative source (i.e., the secondary website) during a task, either because the user did not continue beyond the warning to the primary source or because the user sought to confirm the accuracy of the information from the primary source.\footnote{CTR and AVR are closely related: a non-clickthrough is a type of alternative visit. As a result, $\textrm{AVR} \ge 1 - \textrm{CTR}$.}
We used this data to compute each warning's AVR: the proportion of tasks where a participant visited an alternative source before completing the task.
AVR is a new metric we devised for empirically measuring the behavioral effects of a disinformation warning.\footnote{Another advantage of the AVR metric is that it is available in control conditions, not just treatment conditions. We did not record alternative visits for control rounds in the laboratory study, but we make extensive use of control round AVR  in the crowdworker study.}
A high AVR indicates that a warning can influence users to decide to visit secondary sources of information.\footnote{AVR does not capture user perceptions of warnings or the accuracy of user beliefs, which is why we pair this approach with qualitative methods.
It is an important open question whether encounters with high AVR warnings are associated with more accurate beliefs, easier correction of misperceptions, or better ability to distinguish disinformation from real news.}
In some cases this will cause a user not to see the disinformation at all, and in all cases it exposes the user to alternative information.

\paragraph{Interview}
After the final round, we informed participants of the true purpose of the study, then conducted the interview.
We first asked about the participant's general understanding of disinformation: how they defined disinformation, what made them believe a website might contain disinformation, and if they had ever encountered content that they recognized as disinformation.
Next, we asked the participant to describe their reactions to the warnings that they encountered during the study.
We prompted participants to elaborate on these responses until we could determine whether the participant had noticed and comprehended the warnings (\hyperref[item:labRQ1]{RQ1} and \hyperref[item:labRQ2]{RQ2}).

Before the next round of questions, we showed the participant printouts of the contextual and interstitial warnings used in the study.
We then asked whether the participant believed the warnings would be effective in use, if they felt that one format would be more effective than the other, and if they had recommendations for how disinformation warnings in general could be made more effective.

Finally, we asked about the participant's demographics, academic background, and level of news consumption.\footnote{We list all survey questions in supporting materials~\cite{disinfowarn-sm}.}

\paragraph{Coding}
We combined interview transcripts with our notes to form a single report for each session, then open coded the reports using Dedoose.
One author performed the initial coding, producing 253 unique codes, then condensed the codes into themes.
A second author validated this work, ensuring that the codes accurately reflected the study data and that the proposed themes were justified by the codes.

\begin{table*}[t]
	\small
	\centering
	\caption{We measured clickthrough rates (CTR) and alternative visit rates (AVR) in treatment rounds of the laboratory study.}
	\label{tbl:laboutcomes}
	\begin{tabular}{|l|l|c|c|c|c|}
		\hline
		\multicolumn{1}{|c|}{\cellcolor[HTML]{FFFFFF}{\color[HTML]{333333} }}                                 & \multicolumn{1}{c|}{\cellcolor[HTML]{FFFFFF}{\color[HTML]{333333} }}                                                                                                 & \multicolumn{2}{c|}{\cellcolor[HTML]{FFFFFF}\textbf{\begin{tabular}[c]{@{}c@{}}Contextual Warning\TBstrut\end{tabular}}} & \multicolumn{2}{c|}{\cellcolor[HTML]{FFFFFF}\textbf{\begin{tabular}[c]{@{}c@{}}Interstitial Warning\TBstrut\end{tabular}}} \\ \cline{3-6}
		\multicolumn{1}{|c|}{\multirow{-2}{*}{\cellcolor[HTML]{FFFFFF}{\color[HTML]{333333} \textbf{Round\Bstrut}}}} & \multicolumn{1}{c|}{\multirow{-2}{*}{\cellcolor[HTML]{FFFFFF}{\color[HTML]{333333} \textbf{Participant Instructions\Bstrut}}}}                                              & \textbf{CTR\TBstrut}                                            & \textbf{AVR\TBstrut}                                           & \textbf{CTR\TBstrut}                                             & \textbf{AVR\TBstrut}                                            \\ \hline
		\textbf{Control 1}                                                                                    & \begin{tabular}[c]{@{}l@{}}Find the total area of Italy in square kilometers\Tstrut\\ on Wikipedia or WorldAtlas.\Bstrut\end{tabular}                                & --                                                              & --                                                             & --                                                               & --                                                              \\ \hline
		\textbf{Control 2}                                                                                    & \begin{tabular}[c]{@{}l@{}}Report the price of a pair of men's New Balance 574\Tstrut\\ on JoesNewBalanceOutlet or 6pm.com.\Bstrut\end{tabular}                      & --                                                              & --                                                             & --                                                               & --                                                              \\ \hline
		\textbf{Treatment 1}                                                                                  & \begin{tabular}[c]{@{}l@{}}Find the political party of Taiwan's Premier on\Tstrut\\ TheNewsLens or FocusTaiwan.\Bstrut\end{tabular}                                  & 15/20                                                           & 7/20                                                           & 7/20                                                             & 13/20                                                           \\ \hline
		\textbf{Treatment 2}                                                                                  & \begin{tabular}[c]{@{}l@{}}Find the name of the girl reported missing in Barbados\Tstrut\\ on Feb 11, 2019 on BarbadosToday or LoopNewsBarbados.\Bstrut\end{tabular} & 18/20                                                           & 4/20                                                           & 11/20                                                            & 10/20                                                           \\ \hline
	\end{tabular}
\end{table*}

\subsection{Participant Recruiting}
We recruited participants through the Princeton University Survey Research Center, which advertises to randomly selected students.
We also sent recruiting emails to distribution lists of various student organizations.
We received 76 responses and selected 40 participants.
Our participant group consisted of 16 women and 24 men aged 18-28 years old, studying across 17 disciplines.

Clearly this sample is biased in several respects, including age, education level, and social group.
Later in this work, we evaluate a significantly more diverse sample recruited online (see \hyperref[item:crowdRQ1]{RQ1} in Section~\ref{sec:mturkstudy}).
In the context of security warning studies, student populations have been shown to provide similar results to more representative samples~\cite{sotirakopoulos2011challenges}.

The recruiting and consent materials provided to participants indicated that the study would take 30-45 minutes and focus on the user experience of search engines, with no mention of disinformation or browser warnings.
Participants signed consent forms before beginning the study and were paid \$15.
The study was approved by the Princeton IRB.

\subsection{Results}
We present quantitative results for the warnings' behavioral effects (Table~\ref{tbl:laboutcomes}).
We also discuss how notice and comprehension related to participant behavior and present qualitative results on user opinions and perceptions of the warnings.

\subsubsection{Behavioral Effects}
\paragraph{Contextual}
The CTR for the contextual warning was very high: $33/40$.
There were a total of $11/40$ alternative visits: 7 non-clickthroughs and 4 occasions where a participant who clicked through a warning went back to search again using the secondary source.

\paragraph{Interstitial}
The CTR for the interstitial warning was much lower: $18/40$.
We observed 1 alternative visit after a clickthrough and 22 alternative visits after non-clickthroughs, for a total AVR of $23/40$.

\subsubsection{Notice and Comprehension}
\paragraph{Contextual}
In the contextual treatment group, 13 out of 20 participants stated during interviews that they were not aware they had been shown disinformation warnings.
All of these participants clicked through both warnings.
4 reported that they did not notice the warnings at all. Among the 16 that did notice the warnings, 9 noticed the icons but not the text.

\paragraph{Interstitial}
All 20 participants noticed the interstitial warnings. 12 understood that the warnings were about disinformation.
7 believed the warnings communicated a risk of ``harm,'' a ``virus,'' or another ``security threat'' and quickly clicked to go back without reading the text.
The remaining participant clicked through both warnings; when asked why, he explained that he was focused on completing the study and ``probably would have reacted differently'' outside of the study.

\subsubsection{Opinions on Disinformation Warnings}
\label{subsubsec:opinions}
As part of the interview, we displayed printouts of both warning designs and asked for the participant's opinions about the warnings' relative merits and general effectiveness.

\paragraph{Contextual}
When asked which warning design they believed would be more effective in general, a small minority (6/40) chose the contextual warning.
5 of these participants were in the interstitial treatment group.

5 participants noted that the contextual warning could be seen before a user ``commits'' by clicking a link.
1 participant explained, ``\textit{you're immediately presented with alternatives, whereas for the interstitial I'm already there and committed a click, so I want to go forward.}''
Another preferred the contextual warning because it was easier to bypass: ``\textit{[I] just wanted to find a link to click on very quickly, it doesn't take as much effort to avoid compared to the interstitial.}''

5 other participants emphasized the ``always-on'' nature of the contextual warning.
1 participant liked how they could ``\textit{always see the warning when browsing Google search results without having to click around.}''
Another felt that the contextual warning was paternalistic because it tilted the otherwise level playing field among search results, ``\textit{direct[ing] you to which [results] you should visit}.''

15 participants said that the contextual warning was not noticeable.
1 specified that ``\textit{the exclamation point is very subtle... you're not going to notice it}.''

\paragraph{Interstitial}
34 participants---an overwhelming majority---believed that the interstitial warning would be more effective in general.
When asked why, 32 mentioned that it was more noticeable.
17 mentioned the color red, with 1 participant noting that ``\textit{everybody knows red means stop}.''

19 participants remarked on how the the warning requires user input to proceed.
1 participant observed that ``\textit{it stops the flow of the user and forces them to take an action}.''
Other responses suggest that design cues contributed to the warning's effectiveness; participants mentioned that the red color and text ``\textit{implied that the user is in danger}'' and that the text was ``\textit{more instructive than the text on the contextual warning}.''

When asked about drawbacks to the interstitial warning, 5 participants focused on the inconvenience and the potential for warning fatigue.
1 participant noted that they would ``\textit{probably turn it off in the settings}'' if the warning showed up frequently.

\paragraph{Improving Warning Designs}
Many participants (17) suggested that more informative disinformation warnings would be more effective.
Recommendations included adding ``\textit{more about why this particular site was flagged},'' definitions of terms, and more explicit descriptions of potential harms.
Conversely, 7 others urged ``\textit{short and concise}'' messages that ``\textit{[get] the point across quickly}.''

5 participants suggested using different warnings depending on severity and whether the user had visited the website before.
Another 5 recommended that warnings persist even if a user had visited a website before, arguing that warnings would be more effective if users were ``\textit{consistently reminded that [the page] may not be completely safe or factual.}''

Participants also suggested alternate warning forms: popups, banner messages, or highlighting false claims.

\paragraph{Trust}
The source of the warning was important to many participants.
8 indicated that they were more likely to be deterred by a browser warning if they knew that it was triggered by Google, since they trusted the company's judgment.
1 participant explained that they clicked through the interstitial warning because they understood that Princeton University had flagged the website, and they felt that the university was not a credible source of judgment about online disinformation.
Another theme underlying trust judgments was previous experiences with browser warnings.
7 participants expressed that they distrusted the warnings due to previous encounters with false positive warnings or overly restrictive site blockers on institutional networks (e.g., in high school).

\paragraph{Risk}
7 participants expressed the opinion that disinformation is not a serious threat or that it is not as harmful as malware.
One participant explained that they typically comply with browser warnings but reacted differently to the disinformation warning: ``\textit{I don't like the idea of someone telling me where or what I am allowed to access. You can give me suggestions. It was because I realized it was a disinformation warning and not a malware warning that I went back to try and get to the website.}''
Another participant characterized this sentiment sharply, saying ``\textit{[d]isinformation warnings should not make it harder to access the site}.''

\subsection{Discussion}
\paragraph{Contextual vs. Interstitial Warnings}
The interstitial warning was distinctly more noticeable and comprehensible than the contextual warning, and also far more effective at inducing behavioral changes.
Similar findings in security warning research prompted the field to shift from contextual to interstitial formats.
Platforms are only just beginning to make this shift for disinformation; contextual warnings are currently the dominant format in both research and deployment.
While contextual warnings may still have a role in countering disinformation, interstitial warnings and other forms of active, interruptive warnings clearly merit further consideration.

\paragraph{Impact of Visual Design Choices}
The iconography, colors, and text styles in our warnings impacted participant attention, comprehension, and behavior.
The icon we added to the contextual warning made the warning more noticeable but did not necessarily aid with comprehension, as many participants who noticed the icon still failed to notice the text.
The red background of the interstitial warning contributed to its effectiveness, but may also have reduced comprehension as participants seemed to react quickly upon seeing the red color without taking the time to read or understand the warning.
Again drawing from security warning research, future work should use comparative effectiveness studies to isolate the effects of individual visual design choices.

\subsubsection{Mechanisms of Effect}
\label{subsubsec:lab-mechanisms}
So few participants complied with the contextual warning that it is difficult to draw conclusions about what caused the behavioral effect.
For the interstitial warning, however, we found evidence for three different mechanisms by which the warning induced behavioral changes.

\paragraph{Informativeness}
Warning science literature focuses on educating the user and enabling them to make an informed decision about how to proceed~\cite{wogalter2006purposes,stewart1994intended}.
Across both warning designs we tested, participants who understood the warning visited an alternative website in over half of cases (21/38), while participants who did not understand the warning did so in only a third of cases (13/42).
Moreover, nearly half of participants recommended making the warnings more informative to improve effectiveness.
These results reinforce that informing users is a possible mechanism of effect for interstitial disinformation warnings.

\paragraph{Fear of Harm}
The interstitial warnings had a threatening appearance, which many participants identified as a factor in why they did not continue.
Some participants visited an alternative site because they perceived a non-specific risk of personal harm, without comprehending the warning.
Other participants misinterpreted the warning and believed that it described a risk of receiving a computer virus or other security threat.
If a warning conveys a risk of harm, it should be specific and narrowly scoped; otherwise users may perceive the warning as irrelevant or a false positive, which could reduce the behavioral effect~\cite{egelman2013importance,akhawe2013alice,bravolillo2014revisiting,weinberger2016week,vance2018tuning}.
As long as the specific harm is clear, though, using design cues to further convey a general risk of harm may improve warning effectiveness.

\paragraph{Friction}
The interstitial warning's strong effect was due, in part, to the friction it introduced into the task workflow.
Some participants preferred to choose another source rather than read the warning, decide whether to believe and comply with it, and click through it.
As with the ``fear of harm'' mechanism, friction must be carefully calibrated to avoid inducing warning fatigue or habituating users to ignore warnings.
Friction also has serious drawbacks as a causal mechanism: it degrades the user experience, makes platforms more difficult to use, and does not rely on an informed decision about disinformation.

\subsubsection{Limitations}
Security warning research has observed challenges in studying behavioral responses to risks in laboratory settings, particularly with respect to ecological validity~\cite{sotirakopoulos2011challenges}.
We encountered similar challenges in this study.

Our sample was small in size and biased in several ways; our findings should be understood as illustrative but not representative.
It is important to identify if our findings can be replicated by larger, more diverse populations.

Because participants used our computer and we were watching during the task, some participants reported that they behaved differently in the study than they might have in real life.
Others may not have reported this effect because they were reluctant to admit that they behaved with bias or because the effect was unconscious.
An experimental design that allows participants to use their own computers in their own environments (i.e., not in a laboratory) could offer more realistic observations of how participants assess risk.

Although participants appeared to be driven to complete the research tasks, they were not personally invested in completing the tasks or finding correct answers to the queries.
The role-playing aspect of the study may not have been strongly immersive, and there were no extrinsic rewards or penalties that incentivized correct answers.

Finally, because our search tasks did not pertain to participants' social or political contexts, participants had little reason to engage in motivated reasoning.
Motivated reasoning can strongly influence a user's perceptions of information relevant to their social or political beliefs~\cite{kahan2012ideology}, so in those contexts, the warning effects that we demonstrate may be weaker.

\vspace{-1mm}
\section{Crowdworker Study}
\label{sec:mturkstudy}
In our second study, we aimed to verify the behavioral effects of interstitial disinformation warnings. We also examined the mechanisms for those effects, so that we could reason about the utility and limitations of deploying the warnings.

Our research questions were:

\begin{itemize}[wide]
	\item[] \label{item:crowdRQ1} \textbf{RQ1: Do interstitial disinformation warnings cause users to choose alternative sources of information?}
	We investigated whether population sample bias or task design significantly affected the results of our laboratory study. We recruited a larger, more diverse participant pool from Amazon Mechanical Turk (Section~\ref{subsec:mturkrecruiting}) and adjusted the task to account for limitations in the laboratory study (Section~\ref{subsec:mturktask}).
	\item[] \label{item:crowdRQ2} \textbf{RQ2: Do interstitial warnings that effectively inform users about the risks of disinformation cause users to choose alternative sources of information?}
	We tested whether participants understood the warnings, then compared the behavioral effects of informative and uninformative warnings to isolate the impact of informativeness on behavior.
	\item[] \label{item:crowdRQ3} \textbf{RQ3: Do interstitial warnings that communicate a risk of personal harm cause users to choose alternative sources of information? }
	We tested whether warnings caused participants to fear harm, then compared the behavioral effects of warnings that did and did not evoke a fear of harm.
	\item[] \label{item:crowdRQ4} \textbf{RQ4: Does user partisanship (with respect to U.S. politics) moderate behavioral effects or perceptions of interstitial warnings?}
	Research in political science indicates that political orientation affects judgments of information credibility~\cite{pennycook2019fighting} and efficacy of misinformation warnings~\cite{pennycook2019implied}.
	We included this research question to detect if partisan alignment created a bimodal distribution in responses to warnings.
\end{itemize}

The task structure and key behavioral outcomes remained the same as in the laboratory study.
We informed participants that they were joining a study of search engine usage and research behaviors. We then guided participants through four research tasks using a search engine, alternating between control and treatment rounds.
In each treatment round, the participant encountered one of eight candidate interstitial disinformation warnings after clicking certain search results.
We measured whether the participant clicked through the warning and whether they visited an alternative website.
We examine the CTR and AVR across all observations to answer \hyperref[item:crowdRQ1]{RQ1}.

We used surveys after each warning encounter to measure how informative the warning was and how strongly the participant perceived the warning to convey a risk of harm.
\hyperref[item:crowdRQ2]{RQ2} and \hyperref[item:crowdRQ3]{RQ3} concern the relationship between AVR and these survey responses.
A standard analysis approach would have been to randomly assign participants to warnings, then compute statistical tests across the conditions.
Unless the differences in effect between warnings were dramatic, however, this approach would have required a massive number of observations on each warning to establish statistical significance.

We instead employed a multi-armed bandit algorithm, which allows efficient exploration of a larger design space than is traditionally possible.
As we received successive observations, the bandit increased the odds that participants encountered the warnings proving to be most and least informative and most and least effective at conveying fear of harm.
After all observations were completed, significantly more participants had encountered these top- and bottom-performing warnings, providing us with the statistical power needed to test our hypotheses.
Section~\ref{subsubsec:bandit} discusses the design and implementation of the multi-armed bandit algorithm.

\subsection{Warning Designs}
\label{subsec:mturkwarnings}
We created eight candidate interstitial disinformation warnings: four designed for informativeness and four designed to evoke fear of harm (Table~\ref{tbl:mturkwarnings}).
The warnings shared a common layout, consisting of an icon, a title, a primary message, an optional detailed message, and two buttons.
This layout differed from the laboratory interstitial warning in two ways.

First, in the laboratory warning design, the detailed message (and the ``Continue'' button) were hidden at first and would only be revealed after the participant clicked ``Learn more.''
In the crowdworker study, we wanted to ensure that the full warning message was always displayed, because part of what we were measuring was the effect of different messages.
We eliminated the ``Learn more'' button and instead displayed the detailed message and the ``Continue'' button on all warnings.

The second change addressed the ``Back to safety'' button.
This button text implied that the user was in danger, which was inappropriate for the informative warnings.
We changed the button to read ``Go back'' and applied this change to both informative and harm-based warnings in order to maintain a common interaction design across all warnings.

For both groups of warnings, we generated several options for icons, titles, and messages. We then created candidate designs by choosing combinations of these elements and inserting them into the layout template.

\begin{figure*}[!htb]
	\minipage[t]{0.5\textwidth}
	\centering
	\framebox{\includegraphics[width=.9\linewidth]{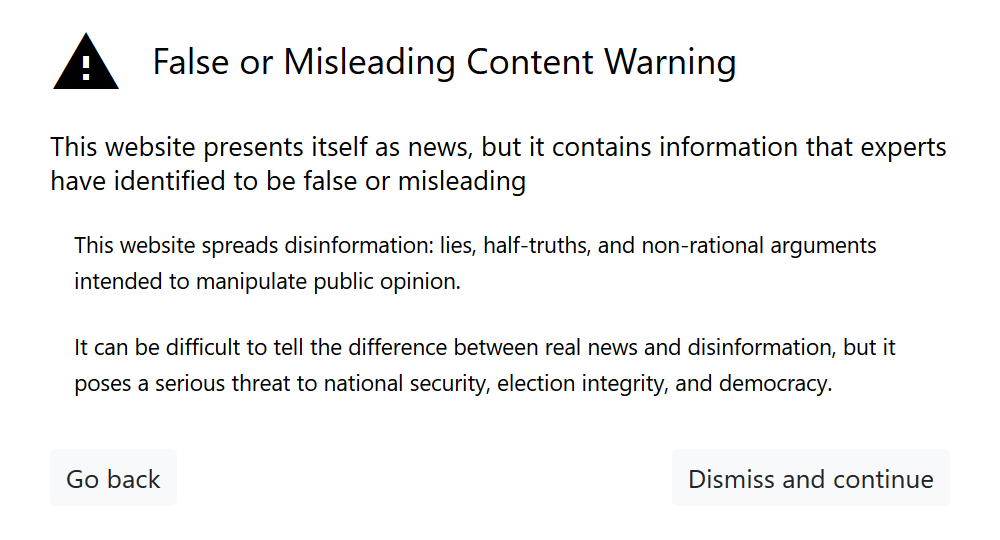}}
	\subcaption{Informative warning design \textit{i2}}\label{fig:warningi2}
	\endminipage\hfill
	\minipage[t]{0.5\textwidth}
	\centering
	\framebox{\includegraphics[width=.9\linewidth]{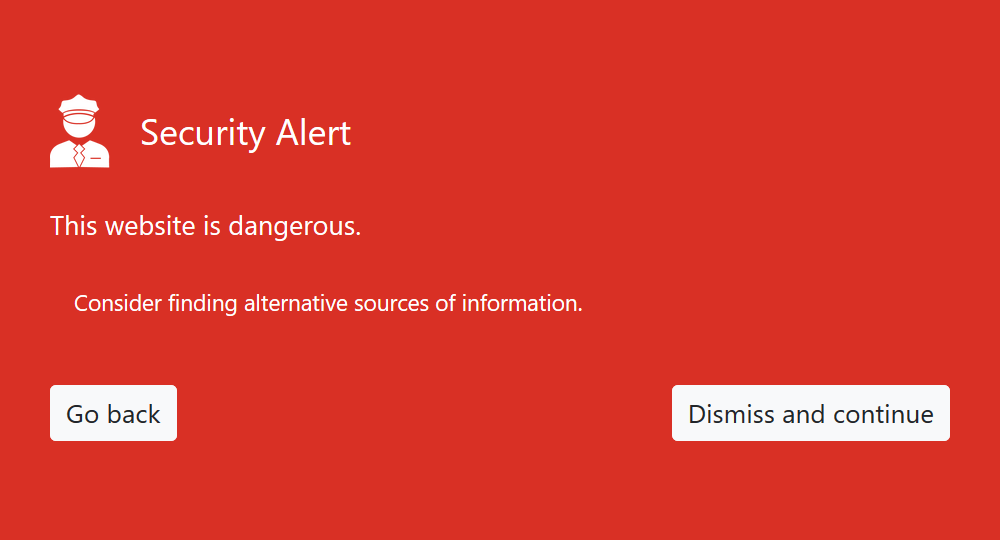}}
	\subcaption{Harm warning design \textit{h3}}\label{fig:warningh3}
	\endminipage
	\caption{Examples of warning designs in the crowdworker study. Table~\ref{tbl:mturkwarnings} describes all eight warning designs.}
	\label{fig:mturkwarnings}
\end{figure*}

\paragraph{Informative Warnings}
We designed the informative warnings to be visually nonthreatening and clearly explanatory in their messages (see Figure~\ref{fig:warningi2}).
The warnings included one of two icons---either a generic exclamation point icon or a policeman silhouette---and displayed black text against a white background.
The warning messages explained the nature and consequences of disinformation in varying detail: some explicitly defined the term ``disinformation,'' some asserted that ``experts'' had labeled the site as containing ``false or misleading'' information, and others provided clear guidance on how to behave (``finding alternative sources of information'').

\paragraph{Harm Warnings}
The harm warnings contained less text and used forceful language, colors, and icons to suggest a serious threat (see Figure~\ref{fig:warningh3}).
The warnings used either a skull-and-crossbones icon or a policeman icon, and the content was colored white against a red background.
The most extreme warning design simply said: ``WARNING: This website is dangerous.''
The other three warning designs were titled ``Security Alert'' and indicated in their messages that the threat had to do with information quality.

\addtolength{\tabcolsep}{-3pt} 
\begin{table*}[!htb]
	\centering
	\caption{We developed eight interstitial warning designs for the crowdworker study. Figure~\ref{fig:mturkwarnings} shows sample designs.}
	\label{tbl:mturkwarnings}
	\scalebox{0.8}{
		\begin{tabular}{|l|l|l|l|l|l|l|l|l|}
			\hline
			\textbf{}                                                          & \multicolumn{4}{c|}{\textbf{Harm (white on red background)\TBstrut}}                                                                                                                                                                                                                                                                                                                                                                                                                                  & \multicolumn{4}{c|}{\textbf{Informativeness (black on white background)\TBstrut}}                                                                                                                                                                                                                                                                                                                                                                                                                                                                                                                                                                                                                                                                                                                                                                                                                                                                                                                                                                                                                                   \\ \hline
			\multicolumn{1}{|c|}{\textbf{ID}}                                  & \multicolumn{1}{c|}{\textbf{h1}}                                                                                   & \multicolumn{1}{c|}{\textbf{h2}}                                                                                   & \multicolumn{1}{c|}{\textbf{h3}}                                                                                             & \multicolumn{1}{c|}{\textbf{h4}}                                                                                             & \multicolumn{1}{c|}{\textbf{i1}}                                                                                                                                                                                                                                                                                                                                                         & \multicolumn{1}{c|}{\textbf{i2}}                                                                                                                                                                                                                                                                                                                                                         & \multicolumn{1}{c|}{\textbf{i3}}                                                                                             & \multicolumn{1}{c|}{\textbf{i4\TBstrut}}                                                                                                                                                       \\ \hline
			\textbf{Icon}                                                      & \multicolumn{1}{c|}{\begin{tabular}[t]{@{}c@{}}Skull\\ \includegraphics[width=.2in]{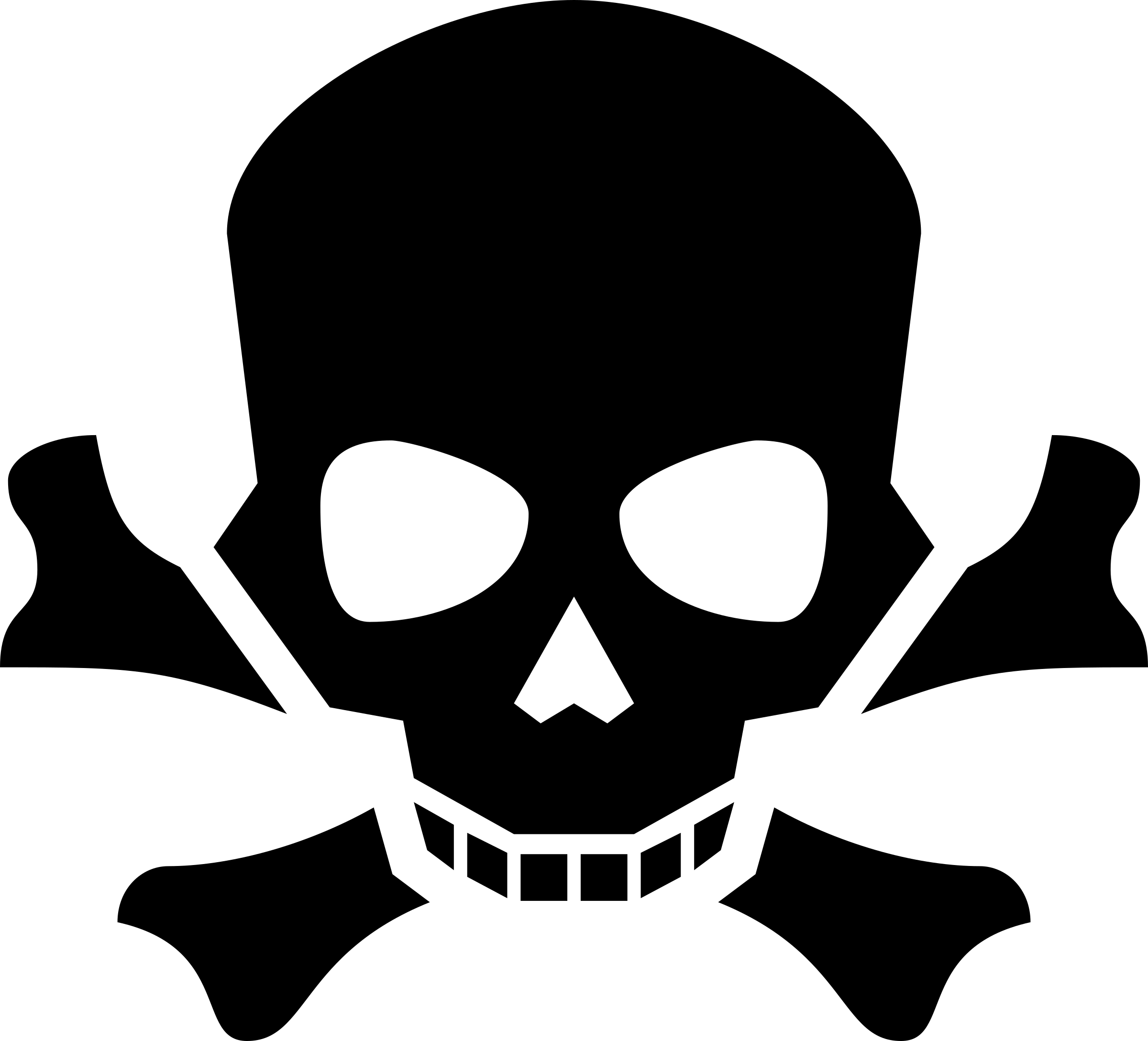}\end{tabular}} & \multicolumn{1}{c|}{\begin{tabular}[t]{@{}c@{}}Skull\\ \includegraphics[width=.2in]{skull_black.png}\end{tabular}} & \multicolumn{1}{c|}{\begin{tabular}[t]{@{}c@{}}Policeman\Tstrut\\ \includegraphics[width=.2in]{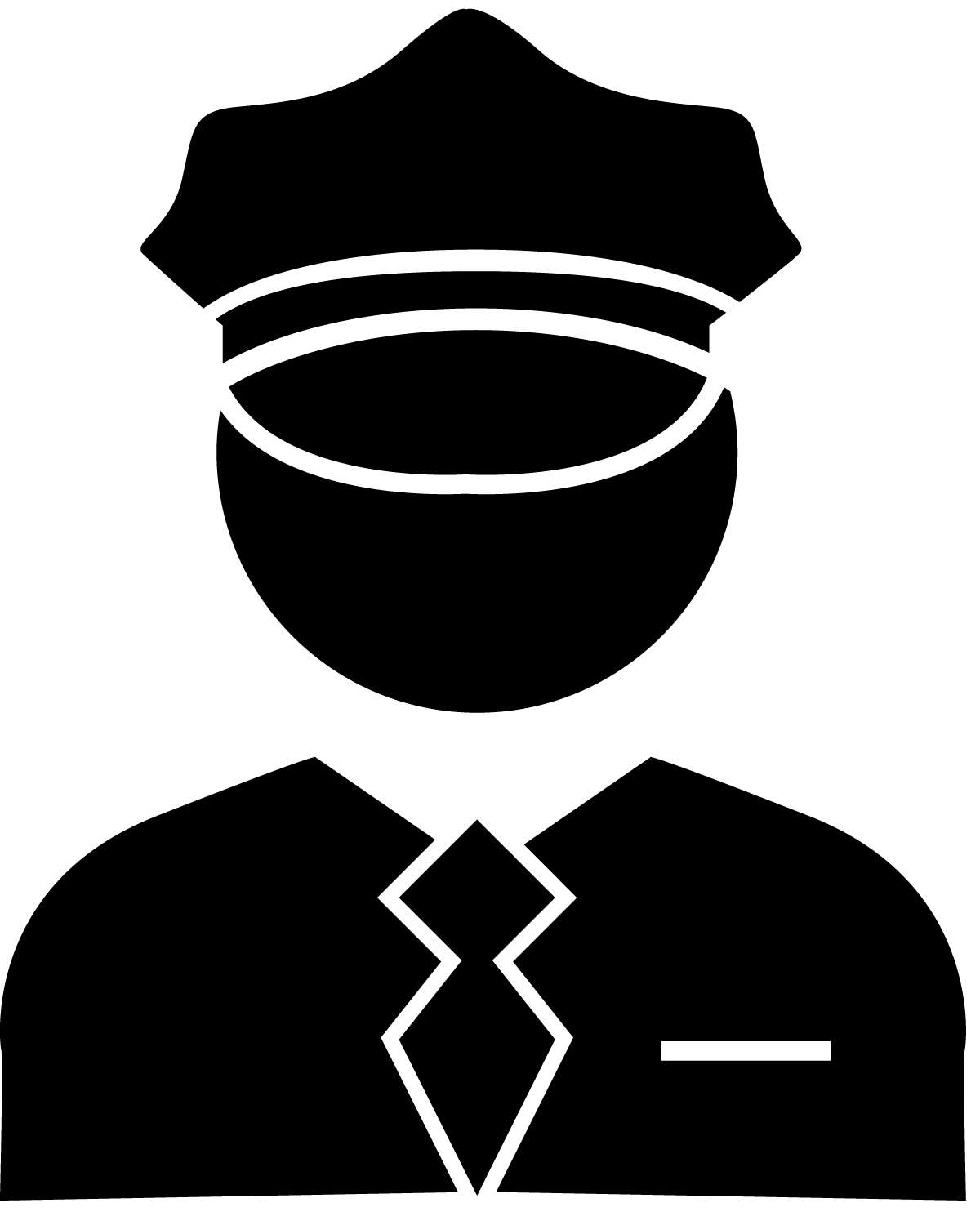}\end{tabular}} & \multicolumn{1}{c|}{\begin{tabular}[t]{@{}c@{}}Policeman\Tstrut\\ \includegraphics[width=.2in]{pman_black.png}\end{tabular}} & \multicolumn{1}{c|}{\begin{tabular}[t]{@{}c@{}}Exclamation\\ \includegraphics[width=.2in]{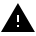}\end{tabular}}                                                                                                                                                                                                                                                               & \multicolumn{1}{c|}{\begin{tabular}[t]{@{}c@{}}Policeman\Tstrut\\ \includegraphics[width=.2in]{pman_black.png}\end{tabular}}                                                                                                                                                                                                                                                             & \multicolumn{1}{c|}{\begin{tabular}[t]{@{}c@{}}Policeman\Tstrut\\ \includegraphics[width=.2in]{pman_black.png}\end{tabular}} & \multicolumn{1}{c|}{\begin{tabular}[t]{@{}c@{}}Exclamation\\ \includegraphics[width=.2in]{warning_black.png}\end{tabular}}                                                                     \\ \hline
			\textbf{Title}                                                     & WARNING                                                                                                            & Security Alert                                                                                                     & Security Alert                                                                                                               & Security Alert                                                                                                               & \begin{tabular}[t]{@{}l@{}}False or\Tstrut\\ Misleading\\ Content\\ Warning \Bstrut\end{tabular}                                                                                                                                                                                                                                                                                         & Fake News Warning                                                                                                                                                                                                                                                                                                                                                                        & \begin{tabular}[t]{@{}l@{}}False or\Tstrut\\ Misleading\\ Content\\ Warning\Bstrut\end{tabular}                              & Fake News Warning                                                                                                                                                                              \\ \hline
			\textbf{\begin{tabular}[t]{@{}l@{}}Primary\\ message\end{tabular}} & \begin{tabular}[t]{@{}l@{}}This website\\ is dangerous.\end{tabular}                                               & \begin{tabular}[t]{@{}l@{}}This website\\ contains\\ misleading\\ or false\\ information.\end{tabular}             & \begin{tabular}[t]{@{}l@{}}This website\\ is dangerous.\end{tabular}                                                         & \begin{tabular}[t]{@{}l@{}}This website\\ contains\\ misleading\\ or false\\ information.\end{tabular}                       & \begin{tabular}[t]{@{}l@{}}This website\Tstrut\\ presents itself\\ as news, but it\\ contains\\ information\\ that experts\\ have identified\\ to be false or\\ misleading\Bstrut\end{tabular}                                                                                                                                                                                           & \begin{tabular}[t]{@{}l@{}}This website\\ contains\\ misleading\\ or false\\ information.\end{tabular}                                                                                                                                                                                                                                                                                   & \begin{tabular}[t]{@{}l@{}}This website\\ contains\\ misleading\\ or false\\ information.\end{tabular}                       & \begin{tabular}[t]{@{}l@{}}This website\Tstrut\\ presents itself\\ as news, but it\\ contains\\ information\\ that experts\\ have identified\\ to be false or\\ misleading\Bstrut\end{tabular} \\ \hline
			\textbf{Details}                                                   & None                                                                                                               & None                                                                                                               & \begin{tabular}[t]{@{}l@{}}Consider\\ finding\\ alternative\\ sources of\\ information.\end{tabular}                         & None                                                                                                                         & \begin{tabular}[t]{@{}l@{}}This website spreads\Tstrut\\ disinformation: lies,\\ half-truths, and\\ non-rational\\ arguments intended\\ to manipulate public\\ opinion.\\ \\ It can be difficult to\\ tell the difference\\ between real news\\ and disinformation,\\ but it poses a serious\\ threat to national\\ security, election\\ integrity, and\\ democracy.\Bstrut\end{tabular} & \begin{tabular}[t]{@{}l@{}}This website spreads\Tstrut\\ disinformation: lies,\\ half-truths, and\\ non-rational\\ arguments intended\\ to manipulate public\\ opinion.\\ \\ It can be difficult to\\ tell the difference\\ between real news\\ and disinformation,\\ but it poses a serious\\ threat to national\\ security, election\\ integrity, and\\ democracy.\Bstrut\end{tabular} & \begin{tabular}[t]{@{}l@{}}Consider\\ finding\\ alternative\\ sources of\\ information.\end{tabular}                         & \begin{tabular}[t]{@{}l@{}}Consider\\ finding\\ alternative\\ sources of\\ information.\end{tabular}                                                                                           \\ \hline
		\end{tabular}
	}
\end{table*}
\addtolength{\tabcolsep}{3pt} 

\begingroup
\addtolength{\tabcolsep}{-2pt} 

\begin{table*}[th!]
	\centering
	\small
	\caption{The crowdworker study involved two control rounds and two treatment rounds. We report outcomes in Table~\ref{tbl:mturkresults}.}
	\label{tbl:mturk-search-tasks}
	\begin{tabular}{|l|l|}
		\hline
		\multicolumn{1}{|c|}{\textbf{Round}} & \multicolumn{1}{c|}{\textbf{Participant Instructions\TBstrut}}                                                                                                                                                                                                                    \\ \hline
		\textbf{Control 1}                   & \begin{tabular}[c]{@{}l@{}}A train from the Massachusetts Bay Transportation Authority derailed in Boston on June 8.\Tstrut\\ Please find the color of the line that the train was on using the query ``MBTA derailment.''\Bstrut\end{tabular}                                       \\ \hline
		\textbf{Control 2}                   & \begin{tabular}[c]{@{}l@{}}In Alabama, the Muscogee tribe filed a lawsuit over a casino that was built on sacred ground.\Tstrut\\ Please find the name of the casino using the query ``muscogee nation casino lawsuit.''\Bstrut\end{tabular}                                         \\ \hline
		\textbf{Treatment 1}                 & \begin{tabular}[c]{@{}l@{}}Please find the names of the two Clatskanie School Board members who were recalled in\Tstrut\\ May 2018 using the query ``clatskanie school board recall.''\Bstrut\end{tabular}                                                                           \\ \hline
		\textbf{Treatment 2}                 & \begin{tabular}[c]{@{}l@{}}Mike Sanders, the Executive of Jackson County, Kansas, was sentenced to prison in 2018\Tstrut\\ for embezzlement. Please find the name of the political party to which he belonged using\\ the query ``mike sanders jackson county.''\Bstrut\end{tabular} \\ \hline
	\end{tabular}
\end{table*}

\endgroup

\subsection{Task Design}
\label{subsec:mturktask}
We aimed to keep procedures for the crowdworker study as similar to the laboratory study as possible, but the different setting and research questions necessitated three changes.

\begin{figure*}[!htb]
	\minipage[t]{0.5\textwidth}
	\centering
	\framebox{\includegraphics[width=.9\linewidth]{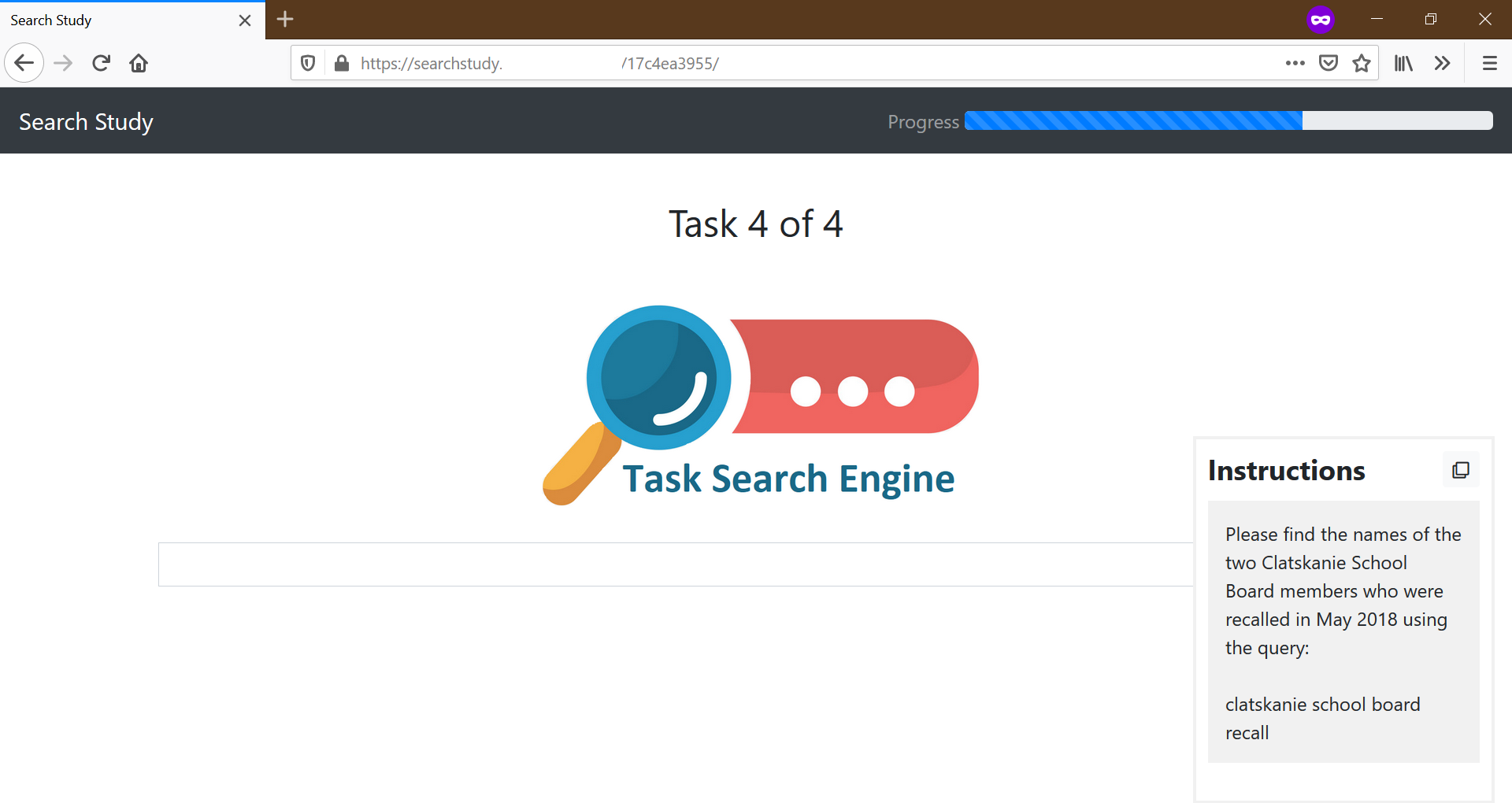}}
	\caption{Each round of the crowdworker study began on this search page.}\label{fig:searchpage}
	\endminipage\hfill
	\minipage[t]{0.5\textwidth}
	\centering
	\framebox{\includegraphics[width=.9\linewidth]{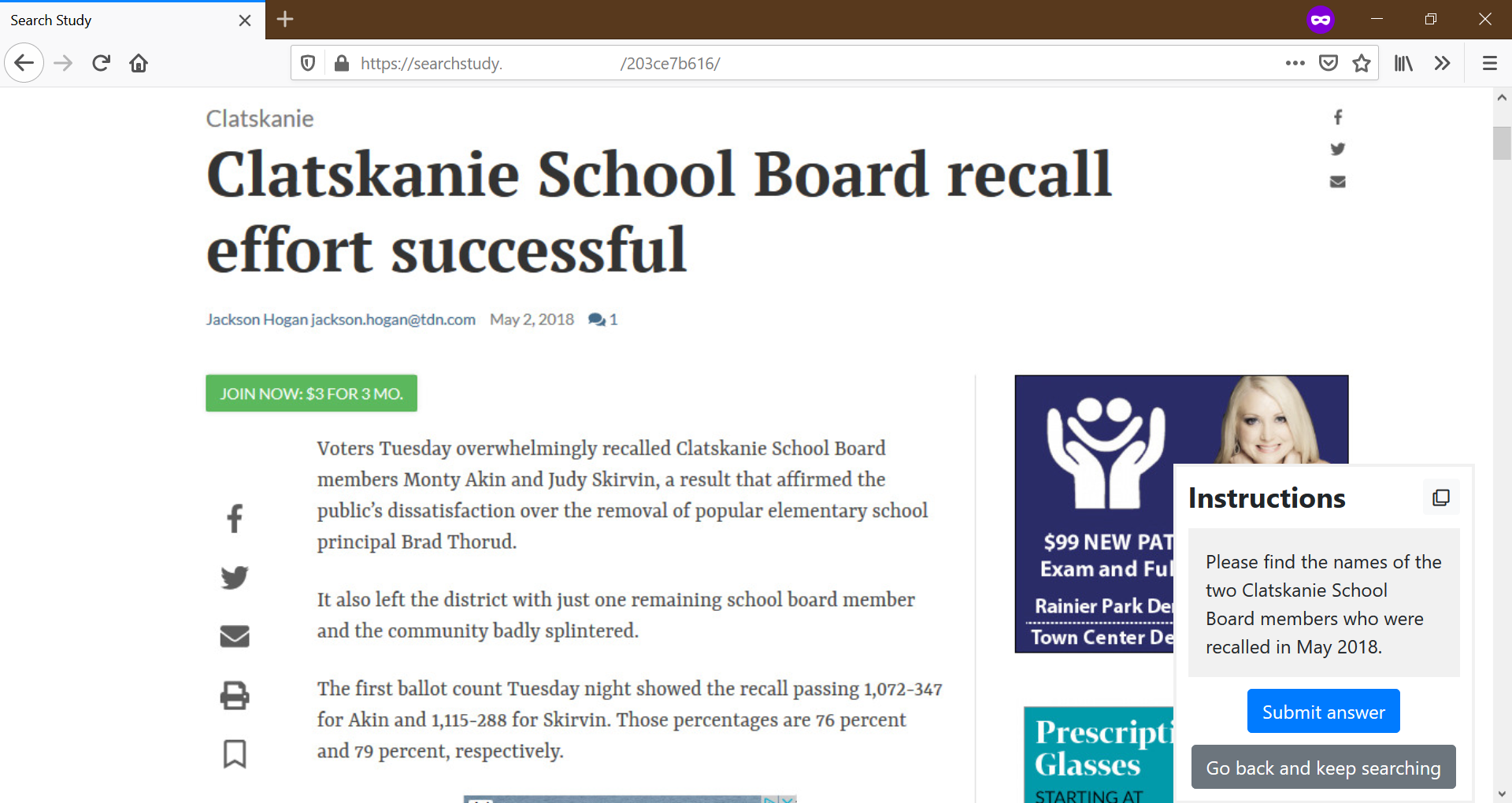}}
	\caption{Clicking on a search result led the participant to a story page containing a screenshot of a real news webpage, instructions, and buttons to submit an answer or navigate back.}\label{fig:storypage}
	\endminipage
\end{figure*}

First, because crowdworkers participate remotely and use their own computers, we could not easily measure their browsing, insert warnings, or control search queries or results.
Instead of using live Google searches, we developed a web application to guide participants through the experiment and realistically simulate a search engine  (Figure~\ref{fig:searchpage}).
We populated results for each query from real Google results for the query, including the snippet for each search result.
In order to simulate the story page after clicking a search result---and to ensure that participants saw the same content---we used full-page screenshots of story pages. Participants could browse these screenshots similar to real webpages (Figure~\ref{fig:storypage}).
Unlike in the laboratory study, we specified the queries to use and did not direct participants to specific sources.

Second, crowdworkers participated in our study to earn a wage.
This is a very different motivation than that of our laboratory subjects.
Crowdworkers may be more focused on completing the task quickly (in order to earn the task fee) than on engaging meaningfully with the study and behaving as ``good'' research participants.
One way we addressed the risk was to ensure that only workers with track records of submitting quality work participated in the study (see Section~\ref{subsec:mturkrecruiting}).
We also offered a bonus of \$1 (43\% of the base fee) to participants who correctly answered all four search questions.
The bonus incentivized crowdworkers to engage with the tasks, read instructions carefully, seek accurate information, and take disinformation warnings seriously.

Finally, we used a series of surveys in lieu of directly observing participants and conducting interviews.
Each participant completed a pre-task survey about their partisan alignment, surveys in each round about their behavior and perception of the warning, and a post-task demographic survey.\footnote{We provide survey details in supporting materials~\cite{disinfowarn-sm}.}

\paragraph{Search Tasks}
As in the laboratory study, we selected facts for participants to retrieve that were reported by multiple sources and were obscure enough that participants would likely be unfamiliar with the topic or sources.
We also ensured that all search results came from news outlets and that no two results came from the same outlet, giving participants a greater variety of choices for sources of information.
All four tasks pertained to events in the U.S. to make the topics more relevant to our U.S.-based participant population.
We also designed the treatment tasks to cover political scandals, so that participants would find it plausible that news outlets might publish disinformation about these topics.
Table~\ref{tbl:mturk-search-tasks} presents the queries for control and treatment rounds in the study.

\paragraph{Procedures}
After accepting our job on Amazon Mechanical Turk, participants navigated to our study web application.
The landing page displayed instructions and a visual guide for the study user interface, then directed participants to begin the first search round.

Each round consisted of a research task where the participant used our simulated search engine to find a particular fact.
The participant began on a generic search query page, which specified the fact to search for and the query to use (Figure~\ref{fig:searchpage}).
When the participant submitted the query, our study application presented a search results page populated with eight results.
Clicking on a result led to a story page containing the news article that the search result snippet described (Figure~\ref{fig:storypage}).
These story pages were full-size screenshots of real news article webpages, allowing participants to scroll though and read the articles as if they were browsing the real webpages.

Three of the results in each round were \textit{target results}: the search result snippets clearly pertained to the query topic, and the story pages contained articles that clearly provided the answer to the task question.
The other five results were \textit{nontarget results}, which linked to story pages that did not contain the answer.
Some nontarget results could be readily identified as irrelevant from the results page snippet (i.e., they contained some search terms but in a different context), while other nontarget results were germane to the query topic but the story page did not contain the answer to the task question.
We ordered search results so that the target results would be easy to find: the top result was always a target result, and the other two target results appeared randomly within the top five results. The rest of the results appeared in random order.

On each story page, an instruction box repeated the question and allowed submitting an answer or returning to the search results.
If the participant returned to the results, each result they previously clicked was grayed out and disabled.

In treatment rounds, the participant saw an interstitial warning the first time they clicked a target result.
Our multi-armed bandit algorithm (Section~\ref{subsubsec:bandit}) selected the warning design that the participant encountered.
All warnings included two buttons: ``Go back'' or ``Dismiss and continue.''
If the user clicked ``Go back'' or the browser back button, they returned to the search results.
The user did not encounter a warning when they clicked on any other result.
If the user clicked ``Dismiss and continue,'' they were taken to the story page, where they could submit an answer or return to the search results.

When the participant submitted an answer, they were presented with a survey about why they chose particular search results.
This survey was a misdirection to maintain the false premise that the experiment was studying search engine usage.
In control rounds, submitting the survey led to the next round.
In treatment rounds, the next page was a second survey, designed to capture whether the participant comprehended the purpose of the warning and whether the participant perceived a risk of harm (see Section~\ref{subsubsec:mturkperceptionmeasures}).
This survey also included an attention check question so that we could discard responses from participants who did not carefully read the instructions.

After completing all four rounds, participants were navigated to a final demographic survey, then compensated.

\subsection{Measuring Participant Perceptions}
\label{subsubsec:mturkperceptionmeasures}
After each treatment round, we presented a survey to measure whether the participant comprehended the warning or perceived a risk of harm.
We developed the survey questions based on our laboratory results and small-scale pilot studies.

\paragraph{Informativeness}
We designed three survey questions to measure whether participants comprehended the purpose of a warning.
Recall that in our laboratory study, participants who misunderstood a warning typically believed the warning was related to malware or another security threat.
We grounded our informativeness questions in this observation, asking participants to indicate whether the warning was about three topics: malware (incorrect), information theft (incorrect), and disinformation (correct).
The survey presented the following statements to participants and asked about agreement on a 5-point Likert scale (``Strongly disagree'' to ``Strongly agree'').

\begin{itemize}
	\item $info_1$: The warning said that this site is trying to install malware.
	\item $info_2$: The warning said that this site is trying to steal my information.
	\item $info_3$: The warning said that this site contains false or misleading information.
\end{itemize}

We used the survey responses to compute an informativeness score $i_{p,w}$ in the range $[{-2},2]$, which captured participant $p$'s certainty that warning $w$ was about disinformation.
$i_{p,w} = 2$ if $p$ ``strongly agreed'' that $w$ was about false or misleading information and ``strongly disagreed'' that $w$ was about malware and stealing information.
For each point deviation from these ``correct'' responses on the Likert scales, we reduced $i_{p,w}$ by 1, resulting in a lower score when the participant was uncertain in their answer or had an incorrect understanding of the warning.
The scoring formula was:

\vspace{-3mm}
\[i_{p,w} = max({-2},\ info_3 - info_2 - info_1 - 1)\]

\paragraph{Harm}
We designed two survey questions to measure whether a warning caused a participant to perceive a risk of harm.
The survey asked about agreement with the following statements, using the same 5-point Likert scale as above.

\begin{itemize}
	\item $harm_1$: After seeing the warning, I believe the website may harm me if I visit it.
	\item $harm_2$: After seeing the warning, I believe the website may harm other people who visit it.
\end{itemize}

The two questions distinguish between personal harm (the first question) and societal harm (the second question). Recall that in our laboratory study, we identified personal harm as a possible mechanism of effect for disinformation warnings. We found when piloting our crowdworker study that participants routinely conflated personal and societal harm when answering survey questions. We expressly asked about these two types of harm to ensure clarity for participants, and we solely used the personal harm response in our analysis. We computed the harm score $h_{p,w}$ for participant $p$ and warning $w$ by projecting their $harm_1$ response into the range $[{-2},2]$:

\vspace{-3mm}
\[h_{p,w} = harm_1 - 3\]

\subsection{Measuring Participant Behaviors}
We measured the same behavioral outcomes as in the laboratory study (Section~\ref{subsec:lab-data}): clickthrough rate (CTR) and alternative visit rate (AVR).
CTR represents the proportion of warning encounters where the participant clicked ``Dismiss and continue.''
AVR measures how often participants clicked on more than one source before submitting an answer. We recorded an alternative visit in a control round when the participant visited more than one story page, and we recorded an alternative visit in a treatment round when the participant visited a different story page after encountering a warning (regardless of whether the the participant clicked through). Measuring AVR in control and treatment rounds enables us to estimate the warning's effect with respect to a base rate.

\subsection{Assigning Warnings}
\label{subsubsec:bandit}
In order to answer our research questions about mechanisms of effect (\hyperref[item:crowdRQ2]{RQ2} and \hyperref[item:crowdRQ3]{RQ3}), we measured for differences in behavioral effects between the warnings that achieved the highest and lowest mean scores for informativeness and harm.

The standard method for comparing effect sizes between treatments is a randomized controlled trial, in which participants are randomly assigned to treatments (often in equal numbers).
The key variable determining how many observations are needed is the estimated difference in effect size between treatments.
If this difference is small, the study will require a large number of observations for each treatment to achieve statistically significant confidence.

When designing our study, we observed that the difference in effect sizes between treatments could be small, meaning that a large sample size could be necessary to evaluate our hypotheses.
Our observations were expensive, however, and although we were testing 16 different conditions (8 warning treatments with 2 score outcomes each), we were only interested in comparing the effects of 4 conditions (with the top and bottom mean scores for each outcome).
We therefore sought a method to efficiently assign participants to warnings so that the top- and bottom-scoring warnings achieved high confidence levels, but the other warnings (which we would not use in our hypothesis tests) would receive fewer observations and therefore consume fewer experimental resources.

In this study, we used an adaptive bandit algorithm to assign participants to warnings based on observations of previous participants.
With each new observation, bandit algorithms update the probability of each condition in a study according to some \textit{reward function} that aligns with the researchers' scientific goals~\cite{lattimore2020bandit}.
Bandits have been widely used in clinical trials, software design optimization, and political opinion research~\cite{hu2006theory, white2012bandit, offer-westort2019adaptive}.
We discuss the full details of our multi-armed bandit implementation in supporting materials~\cite{disinfowarn-sm}.

The reward function in our adaptive experiment preferred disinformation warnings that achieved high and low mean scores for informativeness and harm.
For the first $n=80$ participants, the algorithm assigned all warnings equally.
For the remaining participants, the reward function prioritized warnings with the highest and lowest mean scores for informativeness and harm. As the bandit algorithm iterated, maximum and minimum scoring warnings emerged, and the algorithm improved our confidence in the mean scores for these warnings by prioritizing them for presentation to participants.

\subsection{Participant Recruiting}
\label{subsec:mturkrecruiting}
We collected data from 250 Amazon Mechanical Turk workers who were based in the U.S. and had completed more than 5,000 jobs with an approval rate above 97\%.
We discarded data from 12 workers who failed an attention check question, leaving a sample population of 238.
The population was roughly two-thirds male and over half of participants were between the ages of 30 and 49.
The majority consumed news media at least five days a week and paid somewhat close attention to politics and current events. We provide full population demographics in supporting materials~\cite{disinfowarn-sm}.

Recruiting and consent materials described the study as related to search engine use and did not mention warnings or disinformation. We estimated the total task duration as 15-20 minutes and compensated participants \$2.33 with the opportunity to earn a \$1 bonus (43\%) for retrieving the correct answer for all four queries.\footnote{We expected that nearly all participants would qualify, and 81.5\% did.}
If a participant abandoned the task partway through or exceeded a 2-hour time limit, we discarded their data and recruited a replacement participant.

Our study was approved by the Princeton University IRB.
\subsection{Results}
\begin{figure*}[!htb]
	\centering
  \subfloat{\label{fig:info-outcomes-l}\includegraphics[width=.24\textwidth]{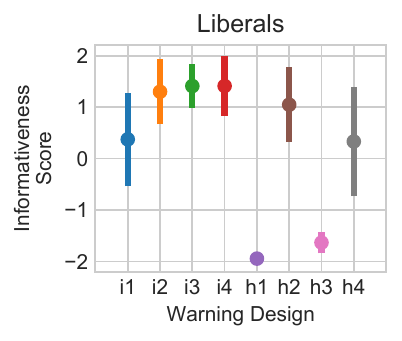}}
  \subfloat{\label{fig:info-outcomes-c}\includegraphics[width=.24\textwidth]{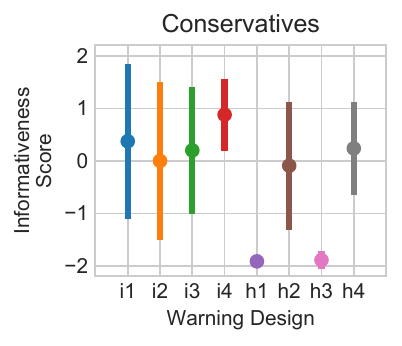}}
  \subfloat{\label{fig:harm-outcomes-l}\includegraphics[width=.24\textwidth]{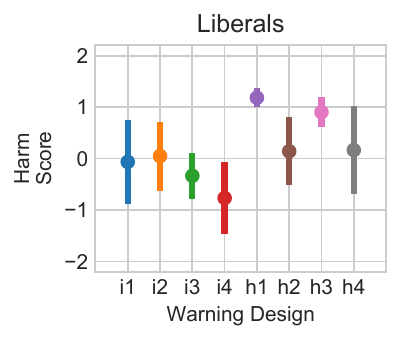}}
	\subfloat{\label{fig:harm-outcomes-c}\includegraphics[width=.24\textwidth]{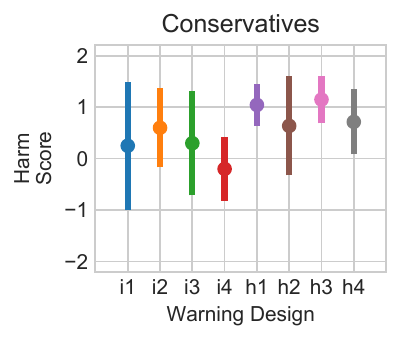}}
	\caption{Mean informativeness and harm scores, with 95\% confidence intervals, for liberals and conservatives.}
	\label{fig:outcomes}
\end{figure*}

We preregistered our analysis methods~\cite{disinfowarn-sm}.
We computed mean ratings and 95\% confidence intervals for informativeness and harm scores (Figure~\ref{fig:outcomes}). For each political alignment and mechanism of effect, we identified the two warnings with the highest and lowest mean scores (Table~\ref{tbl:mturkresults}).\footnote{Our preregistered methods also included a non-overlapping 95\% confidence interval criterion, but it did not affect our warning selection.} We then conducted statistical tests comparing the AVR between these two warnings.\footnote{We did not conduct statistical tests on warning CTRs, because our research questions focused on seeking alternative sources of information.}
We treated clickthroughs and alternative visits as samples drawn from binomial distributions.\footnote{We assumed independence between warning treatments for a participant.}
For tests with large sample sizes, we used a z-test because a normal distribution approximates a binomial distribution; when the sample size was small, we used Fisher's exact test.

\paragraph{Informativeness}
We found that \textit{i3} had a very high mean informativeness score for liberal participants (1.41 on the scale $[-2,2]$).
\textit{i2} and \textit{i4} also had high, consistent informativeness scores for liberals.
As for conservative participants, we found much lower mean informativeness scores for every warning we designed to be informative.
\textit{i4} had the highest mean score (0.88).
The next most informative warning for conservatives was \textit{h4}, which we had intended to convey a risk of harm.

The least informative warning was \textit{h1}, which achieved consistent, extremely low informativeness scores from both liberals and conservatives.
\textit{h1} was the most extreme warning in the harm category; the only text it contained was ``WARNING: This website is dangerous.''

\paragraph{Harm}
\textit{h1} had the highest mean score for evoking fear of harm among liberals (1.18), with high confidence.
We also found that \textit{h1} had a high mean harm score for conservatives, but \textit{h3} had a slightly higher score (1.15).

\textit{i4} had the lowest mean harm score for both political alignments, with a fairly low score among liberals ($-0.76$) and a more neutral score among conservatives ($-0.2$).

\begingroup
\addtolength{\tabcolsep}{-3pt} 

\begin{table}[]
	\small
	\centering
	\caption{We report alternative visit rates (AVR), clickthrough rates (CTR), and mean informativeness ($\bar{i}$) and harm ($\bar{h}$) scores with $95\%$ confidence intervals.}
	\label{tbl:mturkresults}
	\scalebox{0.78}{
		\begin{tabular}{|ll|c|c|c|c|c|c|c|c|c|c|}
			\hline
			\multicolumn{2}{|l|}{}                           & \multicolumn{5}{c|}{\textbf{Liberal\Tstrut}}                                                                                                                                                                                & \multicolumn{5}{c|}{\textbf{Conservative}}                                                                                                                                                                                  \\ \cline{3-12}
			\multicolumn{2}{|l|}{}                           & \textbf{\#\TBstrut} & \textbf{AVR\TBstrut} & \textbf{CTR\TBstrut} & \textbf{$\bm{\bar{i}}$\TBstrut}                                            & \textbf{$\bm{\bar{h}}$\TBstrut}                                            & \textbf{\#\TBstrut} & \textbf{AVR\TBstrut} & \textbf{CTR\TBstrut} & \textbf{$\bm{\bar{i}}$\TBstrut}                                             & \textbf{$\bm{\bar{h}}$\TBstrut}                                           \\ \hline
			\multicolumn{2}{|l|}{\textbf{Control\TBstrut}}   & 318\TBstrut         & 20\%\TBstrut         & --\TBstrut           & --\TBstrut                                                                 & --\TBstrut                                                                 & 158\TBstrut         & 16\%\TBstrut         & --\TBstrut           & --\TBstrut                                                                  & --\TBstrut                                                                \\ \hline
			\multicolumn{2}{|l|}{\textbf{Treatment\TBstrut}} & 318\TBstrut         & 87\%\TBstrut         & 16\%\TBstrut         & --\TBstrut                                                                 & --\TBstrut                                                                 & 158\TBstrut         & 85\%\TBstrut         & 17\%\TBstrut         & --\TBstrut                                                                  & --\TBstrut                                                                \\ \hline
			\multicolumn{12}{|l|}{\textbf{Selected treatments\TBstrut}}                                                                                                                                                                                                                                                                                                                                                                                                                                                  \\ \hline
			& \textbf{h1\TBstrut}              & 120\TBstrut         & 85\%\TBstrut         & 18\%\TBstrut         & \begin{tabular}[c]{@{}c@{}}$-1.94$\Tstrut\\ $\pm 0.06$\Bstrut\end{tabular} & \begin{tabular}[c]{@{}c@{}}$1.18$\Tstrut\\ $\pm 0.18$\Bstrut\end{tabular}  & 46\TBstrut          & 83\%\TBstrut         & 17\%\TBstrut         & \begin{tabular}[c]{@{}c@{}}$-1.91$\Tstrut\\  $\pm 0.11$\Bstrut\end{tabular} & --\TBstrut                                                                \\ \hline
			& \textbf{h3\TBstrut}              & 73\TBstrut          & 84\%\TBstrut         & 18\%\TBstrut         & --\TBstrut                                                                 & --\TBstrut                                                                 & 27\TBstrut          & 81\%\TBstrut         & 22\%\TBstrut         & --\TBstrut                                                                  & \begin{tabular}[c]{@{}c@{}}$1.15$\Tstrut\\ $\pm 0.46$\Bstrut\end{tabular} \\ \hline
			& \textbf{i3\TBstrut}              & 39\TBstrut          & 87\%\TBstrut         & 13\%\TBstrut         & \begin{tabular}[c]{@{}c@{}}$1.41$\Tstrut\\ $\pm 0.43$\Bstrut\end{tabular}  & --\TBstrut                                                                 & 10\TBstrut          & 90\%\TBstrut         & 10\%\TBstrut         & --\TBstrut                                                                  & --\TBstrut                                                                \\ \hline
			& \textbf{i4\TBstrut}              & 17\TBstrut          & 82\%\TBstrut         & 12\%\TBstrut         & --\TBstrut                                                                 & \begin{tabular}[c]{@{}c@{}}$-0.76$\Tstrut\\ $\pm 0.69$\Bstrut\end{tabular} & 25\TBstrut          & 76\%\TBstrut         & 24\%\TBstrut         & \begin{tabular}[c]{@{}c@{}}$0.88$\Tstrut\\ $\pm 0.69$\Bstrut\end{tabular}   & \begin{tabular}[c]{@{}c@{}}$-0.2$\Tstrut\\ $\pm 0.62$\Bstrut\end{tabular} \\ \hline
		\end{tabular}
	}
\end{table}
\endgroup

\paragraph{CTR}
The cumulative CTR across all treatments was 16\%, which was noticeably lower than what we observed for interstitial warnings in the laboratory (40\%).
No individual warning in this study demonstrated a higher CTR than the interstitial warning we tested in the laboratory.

\paragraph{AVR}
The AVR across all treatments was 86\%, compared to 19\% in control rounds.
We used a one-sided z-test to evaluate if this difference was significant, and we found strong support for the hypothesis that the AVR in treatment rounds was greater than the AVR in control rounds ($p = 1.48e^{{-}111}$).

We tested whether there was a significant difference in AVR between the top- and bottom-scoring warnings for informativeness within the liberal and conservative groups.
For liberals, we used a one-sided z-test and failed to reject the null hypothesis that the AVR of the top-scoring warning was less than or equal to the AVR of the bottom-scoring warning ($p = 0.27$).
For conservatives, we used a one-tailed Fisher's exact test (due to the small sample size of conservative participants) and also failed to reject the null hypothesis ($p = 0.54$).

Next, we tested for an AVR difference between the top- and bottom-scoring warnings for harm.
We failed to reject both null hypotheses, with a one-sided z-test for liberals ($p = 0.41$) and a one-tailed Fisher's test for conservatives ($p = 0.74$).

\subsection{Discussion}
\label{subsec:mturkdisc}
We found that interstitial warnings have a strong effect on user behavior, confirming our laboratory study results (\hyperref[item:crowdRQ1]{RQ1}).

The results for warning informativeness were inconclusive (\hyperref[item:crowdRQ2]{RQ2}). We demonstrated that interstitial disinformation warnings can effectively inform users that a website may contain disinformation; we identified warning designs that scored well on average for informing participants. Conveying that a website may contain disinformation can prompt users to think critically about the website's trustworthiness, and critical thinking is an important predictor of a user's ability to correctly judge the accuracy of information~\cite{pennycook2019fighting,pennycook2019lazy}. We did not, however, find evidence that informative warnings have a greater effect on user behavior than uninformative warnings.

The results for warnings conveying a risk of harm were similarly inconclusive (\hyperref[item:crowdRQ3]{RQ3}). We found that warning design can effectively convey a risk of harm, but we did not find evidence that better conveying a risk of harm affects~user~behavior.

We hypothesize that the user experience friction introduced by interstitial warnings may be an important causal factor for changes in user behavior.\footnote{Another possible explanation is a substitution effect in our analysis. The uninformative warnings we examine, for example, could be effective at conveying risk of harm. Future work on the role of friction in disinformation warnings could also shed light on this issue.}
We found evidence for friction as a mechanism of effect in our laboratory study (Section~\ref{subsubsec:lab-mechanisms}), but we did not test the theory in the crowdworker study.

Our results for \hyperref[item:crowdRQ1]{RQ1} and \hyperref[item:crowdRQ2]{RQ2} pose a possible speech dilemma: interstitial disinformation warnings can effectively inform users, but whether a user is informed may have little relation to how they behave in response to warnings. 

Finally, we did not find evidence that partisanship moderates warning perceptions or behaviors (\hyperref[item:crowdRQ4]{RQ4}).
Figure~\ref{fig:outcomes} shows that warning scores were generally similar for liberal and conservative participants, and Table~\ref{tbl:mturkresults} shows that CTRs and AVRs were also close between the groups.

\paragraph{Limitations}
There may be variables we did not measure that moderate the relationships between warning designs, participant perceptions, and behavior.
Detailed qualitative methods, like in our laboratory study, can surface these variables---but are challenging to implement in a crowdworker study.

We also note that while our crowdworker sample was more diverse than our laboratory sample, neither sample was representative of the U.S. population.
The behavioral effects we observed were fairly consistent across demographic groups, though.
Our study population only included individuals located in the U.S.; cross-cultural research is needed to understand if the effects we observed apply globally.

\vspace{-1mm}
\section{Conclusion}
\label{sec:discussion}
In this section, we provide recommendations for future evaluation and deployment of disinformation warnings.

\subsection{Directions for Future Research}
Future work could explore the role of user experience friction in disinformation warnings. We found limited evidence that friction is an important factor in our laboratory study. The results from our crowdworker study also suggest that friction---rather than informativeness or conveying a risk of harm---may be the predominant cause of warning behavioral effects. We did not test friction as a mechanism of effect in our crowdworker study, and our results do not conclusively rule out the mechanisms of effect we did examine. But our results are strongly suggestive, and friction merits further study.

Future work could also evaluate other types of interstitial warnings and interstitial warnings in other contexts, especially social media.
Platforms are already deploying warning popups that a user must dismiss, as well as warning overlays that obscure content until the user clicks~\cite{instagram2019combatting,roth2020updating}.

Another promising direction is evaluating how interstitial warnings interact with factors known to impact warning adherence and receptivity to disinformation.
These factors include repetition of warnings~\cite{egelman2013importance,akhawe2013alice,bravolillo2014revisiting,weinberger2016week,vance2018tuning}, user age and digital literacy~\cite{munger2020effect,guess19less}, user tendency toward cognitive reflection~\cite{tappin2018rethinking,pennycook2019lazy,ross2019beyond}, repeated exposure to inaccurate information~\cite{almuhimedi2014your,pennycook2018prior}, and whether that information aligns with user political preferences~\cite{schaffner2016misinformation,kahan2017misconceptions,guess19less}.

A final direction for future study is exploring possible unintended consequences of interstitial disinformation warnings.
These warnings could create an implied truth effect~\cite{pennycook2019implied}, generally undermine trust in online content~\cite{pennycook2018prior}), cause concern about the warning provider, or lead to warning fatigue~\cite{bravolillo2014revisiting}.

\subsection{Informing Platform Disinformation Warnings}
Interstitial warnings can be effective tools for countering disinformation.
Compared to contextual warnings, interstitial designs are much more noticeable for users and much more capable of informing users about disinformation.
Platforms that use contextual warnings for disinformation should be aware that their warnings may have minimal effects.

Going forward, platforms should follow evidence-based approaches for developing and deploying disinformation warnings.
By conducting internal evaluations, collaborating with independent researchers, and releasing data, platforms can significantly advance their ability to counter disinformation with warnings---just like software vendors have done for over a decade to advance security warnings~\cite{akhawe2013alice,almuhimedi2014your,felt2014experimenting,felt2015improving,weinberger2016week,reeder2018experience}.

\section*{Acknowledgments}
We thank Marshini Chetty and Elissa Redmiles for valuable early feedback on this work. Simone Fischer-H\"{u}bner provided thoughtful shepherding for our paper.

\renewcommand*{\bibfont}{\footnotesize}
\printbibliography

\end{document}